# Coulomb blockade in molecular quantum dots

Kamil Walczak

Institute of Physics, A. Mickiewicz University, Umultowska 85, 61-614 Poznań, Poland
and Instituto de Ciencia de Materiales (CSIC), Cantoblanco, 28049 Madrid, Spain

The rate-equation approach is used to describe sequential tunneling through a molecular junction in the Coulomb blockade regime. Such device is composed of molecular quantum dot (with discrete energy levels) coupled with two metallic electrodes via potential barriers. Based no this model, we calculate nonlinear transport characteristics (conductance-voltage and current-voltage dependences) and compare them with the results obtained within a self-consistent field approach. It is shown that the shape of transport characteristics is determined by the combined effect of the electronic structure of molecular quantum dots and by the Coulomb blockade. In particular, the following phenomena are discussed in detail: the suppression of the current at higher voltages, the charging-induced rectification effect, the charging-generated changes of conductance gap and the temperature-induced as well as broadening-generated smoothing of current steps.

## 1. Introduction

Molecular-scale electronic devices are of growing interests among scientists and engineers due to their potential to become active components for future nanocircuits, namely: interconnects, switches, diodes, transistors, dielectrics, photovoltaics, memories and others [1-4]. Recent advances in experimental techniques made it possible to fabricate such junctions, composed of single molecules (or molecular layers) attached to two (or more) electrodes [5-11]. From an applied point of view, it is desirable to be able to control the way the molecule-to-electrodes couple and eventually connect individual devices into a properly working integrated circuit. On the other hand, for scientific researchers, it is important to understand and appropriately model transport at a molecular scale. Until now, a number of theoretical models have been proposed for calculating transport characteristics through molecular junctions in different transport regimes, using as well parametric as first-principles theories.

      A molecule itself represents quantum dot with discrete energy levels, at least an order of magnitude smaller than semiconductor quantum dots, what is important from the point of view of further miniaturization of electronic devices. Usually molecular contact with the electrodes is suggested to be weak and molecular quantum dot (MQD) can be treated as electrically isolated from metallic electrodes (with a large electron density of states) via potential barriers [5-7]. Under the influence of bias voltage, the current will flow through the

system. However, the number of electrons on the molecular island can change in discrete units by tunneling through potential barriers, but charge is transferred through energetically accessible molecular states (conducting channels). Therefore the shape of transport characteristics is determined by the combined effect of the electronic structure of molecular quantum dots and by Coulomb interactions between electrons.

Let us consider the situation, when one spin-degenerate level is coupled via two tunnel barriers with capacitances $C_1$ and $C_2$ to the source and drain electrodes. The stored electrostatic energy of this capacitor is expressed as: $U = e^2/(2C)$ (where $e$ is an electron charge, while $C = C_1 + C_2$ is total capacitance of the system). This charging energy is usually very small in comparison with the thermal energy $E_T = k_B T$ (where: $k_B$ is Boltzmann constant, while $T$ is an absolute temperature) and Coulomb blockade phenomenon can be fully neglected. In this case the self-consistent field (SCF) method is sufficient to describe transport. However, the capacitance in molecular junctions can reach the value of $C \sim 10^{-19}$ F [6] which corresponds to $U \sim 1$ eV, making single-electron tunneling observable even at room temperature ($E_T \sim 1/40$ eV). Moreover, it is rather clear that the charging energy $U$ is influenced mainly by the extent of the electronic wavefunction. Generally, the more localized wavefunction of the level, the higher the value of the $U$-parameter.

An objective of the present work is to study transport characteristics of metal/MQD/metal junction in the Coulomb blockade (CB) regime within the frames of proposed model. Obtained results will be compared with the results suitable for the case of SCF regime [12]. The origin of Coulomb blockade is the fact that due to small electric capacitance of a small-area junction, its electrostatic energy is changed considerably even when single electron will occupy the molecular island. In our calculations we restrict ourselves to the situation, where internal relaxation processes inside the molecular island are negligible, since relaxation rates are slower than electron transfer rates [13]. Further simplification stems from the assumption that tunnel barriers have resistances much larger than quantum of resistance $h/(2e^2) \approx 12.9$ kΩ (or equivalently conductances much smaller than quantum of conductance $2e^2/h \approx 77.5$ μS). In this case, all the cotunneling processes can be neglected and the current is accurately described by lowest-order perturbation theory. In these conditions, transport can be analyzed in the sequential-tunneling picture using a rate-equation approach [7,14-17]. In other words, the current flowing through the device is a strict consequence of sequential electron transition, where the molecular quantum dot is successively charged and discharged by the tunneling process.

## 2. Description of the model

### 2.1 Energy of the eigenstates

For one spin-degenerate level, the state can be occupied by zero, one or two electrons (with opposite spins). Multi-electron energy for this system can be written as a function of the number of electrons [18]:

$$E(N) = \varepsilon N + \frac{U}{2}(N - N_{eq})^2, \quad (1)$$

The first term of the above Eq.1 is associated with kinetic energy, while the second term is the electrostatic energy stored in the capacitor formed by the conducting level and the electrodes [4]. The electrostatic energy term can be obtained from the relation $Q^2/(2C)$ by noting that $Q = e(N - N_{eq})$ denotes an amount of charge stored in the mentioned capacitor ($N_{eq}$ being the equilibrium number of electrons occupying the particular energy level) and taking into account charging energy defined as $U = e^2/(2C)$. Our simplified treatment neglects electron-electron interactions between electrons occupying different energy levels. This assumption can be adequate in the situation, when separation between particular energy levels is large (~eV).

However, what really matters for transport calculations are energy differences as electrons make transitions between MQD and the electrodes. Such differences are determined independently for every energy level $\varepsilon_i$ taking into consideration its location relatively to Fermi energy $\varepsilon_F$ (or equivalently different values of $N_{eq}$). Here we can distinguish three different situations: (1) $\varepsilon_i < \varepsilon_F$ hence $N_{eq} = 2$ and therefore $\varepsilon_i^+ = E(2) - E(1) = \varepsilon_i - U/2$ and $\varepsilon_i^- = E(1) - E(0) = \varepsilon_i - 3U/2$, (2) $\varepsilon_i = \varepsilon_F$ hence $N_{eq} = 1$ and therefore $\varepsilon_i^+ = E(2) - E(1) = \varepsilon_i + U/2$ and $\varepsilon_i^- = E(1) - E(0) = \varepsilon_i - U/2$, (3) $\varepsilon_i > \varepsilon_F$ hence $N_{eq} = 0$ and therefore $\varepsilon_i^+ = E(2) - E(1) = \varepsilon_i + 3U/2$ and $\varepsilon_i^- = E(1) - E(0) = \varepsilon_i + U/2$.

*2.2 Determination of probabilities*

Every level of the molecular system $i$ has different probabilities to be occupied by an exact number of electrons: zero $P_{0i}$, one $P_{1i}$ or two $P_{2i}$, respectively. All these occupancy probabilities must add up to one (normalization), since on one spin-degenerate level can not be located more than two electrons (due to Pauli exclusion principle), hence:

$$P_{0i} + P_{1i} + P_{2i} = 1. \qquad (2)$$

In the considered situation, we can write down the rate equation that describes four possible transitions between three different states (for details see the Appendix):

$$\frac{d}{dt}\begin{pmatrix} P_{0i} \\ P_{1i} \\ P_{2i} \end{pmatrix} = \begin{bmatrix} -\gamma_{0\to1}^i & \gamma_{1\to0}^i & 0 \\ \gamma_{0\to1}^i & -(\gamma_{1\to0}^i + \gamma_{1\to2}^i) & \gamma_{2\to1}^i \\ 0 & \gamma_{1\to2}^i & -\gamma_{2\to1}^i \end{bmatrix}\begin{pmatrix} P_{0i} \\ P_{1i} \\ P_{2i} \end{pmatrix}. \qquad (3)$$

Here we limit ourselves to steady state conditions, where all the relaxation processes associated with coupling are neglected and therefore all the occupation probabilities are time-independent quantities. Since left hand side of Eq.3 is equal to zero, we have to deal with homogeneous set of three equations. Eliminating the central equation and taking advantage of normalization defined by Eq.2 we obtain:

$$\begin{bmatrix} \gamma_{0\to1}^i & -\gamma_{1\to0}^i & 0 \\ 0 & \gamma_{1\to2}^i & -\gamma_{2\to1}^i \\ 1 & 1 & 1 \end{bmatrix}\begin{pmatrix} P_{0i} \\ P_{1i} \\ P_{2i} \end{pmatrix} = \begin{pmatrix} 0 \\ 0 \\ 1 \end{pmatrix}. \qquad (4)$$

It is easy to solve Eq.4 and express the occupation probabilities with the help of the proper transition rates between analyzed states:

$$P_{0i} = \frac{\gamma_{1\to0}^i \gamma_{2\to1}^i}{\gamma_{1\to0}^i \gamma_{2\to1}^i + \gamma_{0\to1}^i(\gamma_{1\to2}^i + \gamma_{2\to1}^i)}, \qquad (5a)$$

$$P_{1i} = \frac{\gamma^i_{0\to1}\gamma^i_{2\to1}}{\gamma^i_{1\to0}\gamma^i_{2\to1} + \gamma^i_{0\to1}(\gamma^i_{1\to2} + \gamma^i_{2\to1})}, \tag{5b}$$

$$P_{2i} = \frac{\gamma^i_{0\to1}\gamma^i_{1\to2}}{\gamma^i_{1\to0}\gamma^i_{2\to1} + \gamma^i_{0\to1}(\gamma^i_{1\to2} + \gamma^i_{2\to1})}. \tag{5c}$$

*2.3 Transition rates and formula for the current*

To obtain the occupation probabilities from Eqs.5 we have to know the transition rates that can be determined by an assumption of a specific model for the coupling with reservoirs. Using Fermi golden-rule arguments we can express them as follows:

$$\gamma^i_{0\to1} = \frac{2}{\hbar}\int_{-\infty}^{+\infty}d\varepsilon[\Gamma_{1i}f_1 + \Gamma_{2i}f_2]D_i^-, \tag{6a}$$

$$\gamma^i_{1\to0} = \frac{1}{\hbar}\int_{-\infty}^{+\infty}d\varepsilon[\Gamma_{1i}(1-f_1) + \Gamma_{2i}(1-f_2)]D_i^-, \tag{6b}$$

$$\gamma^i_{1\to2} = \frac{1}{\hbar}\int_{-\infty}^{+\infty}d\varepsilon[\Gamma_{1i}f_1 + \Gamma_{2i}f_2]D_i^+, \tag{6c}$$

$$\gamma^i_{2\to1} = \frac{2}{\hbar}\int_{-\infty}^{+\infty}d\varepsilon[\Gamma_{1i}(1-f_1) + \Gamma_{2i}(1-f_2)]D_i^+. \tag{6d}$$

Here $\Gamma_{1i}$ and $\Gamma_{2i}$ are parameters describing the strength of the coupling, while $f_1 = f(\varepsilon - \mu_1)$ and $f_2 = f(\varepsilon - \mu_2)$ are Fermi distribution functions for electrons in source and drain electrodes defined by electrochemical potentials $\mu_1 = \varepsilon_F + eV/2$ and $\mu_2 = \varepsilon_F - eV/2$. Furthermore, $D_i^+$ and $D_i^-$ are densities of states bringing an information about the availability of electrons with energies $\varepsilon_i^+$ and $\varepsilon_i^-$ (defined in subsection 2.1), respectively. Here we assume that such densities have the shape of a Lorentzian curve:

$$D_i^\pm = \frac{1}{\pi}\frac{\Gamma_i}{(\varepsilon - \varepsilon_i^\pm)^2 + \Gamma_i^2}, \tag{7}$$

where: $\Gamma_i = (\Gamma_{1i} + \Gamma_{2i})/2$. However, in the limit of extremely weak coupling $\Gamma_i \to 0$, density of states become Dirac delta functions $D_i^\pm \approx \delta(\varepsilon - \varepsilon_i^\pm)$ and since we limit ourselves to weak-coupling case we will use this approximation throughout this work (except subsection 3.3, where the influence of broadening on transport characteristics will be discussed).

Since every level (conducting channel) contribute independently to the current flowing through the device, general formula for that current can be written as a sum over all accessible energy levels $i$ as (for details see the Appendix):

$$I = \frac{2e}{\hbar}\sum_i\int_{-\infty}^{+\infty}d\varepsilon\Gamma_{1i}\left[P_{2i}(1-f_1)D_i^+ + \frac{P_{1i}}{2}\left[(1-f_1)D_i^- - f_1 D_i^+\right] - P_{0i}f_1 D_i^-\right]. \tag{8}$$

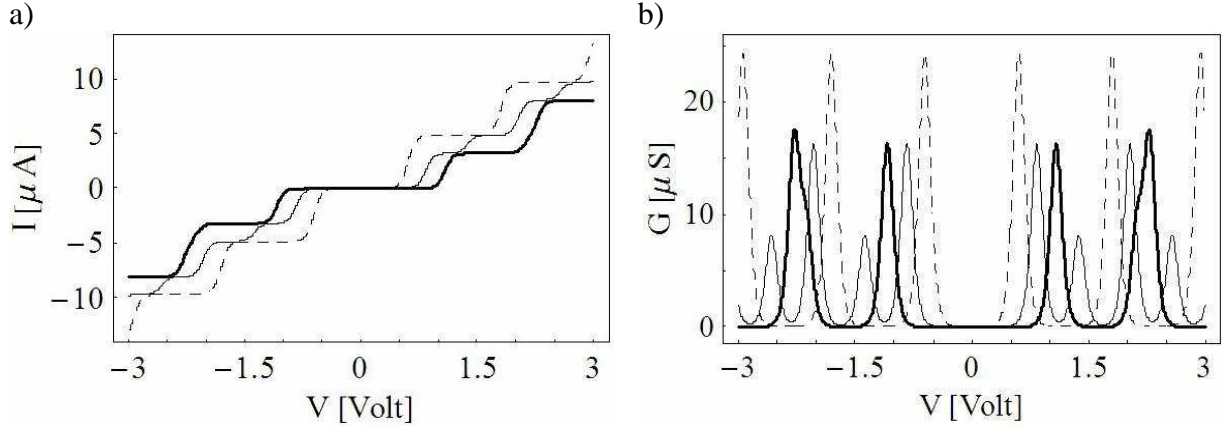

Figure 1: Results for linear chain of 12 carbon atoms for symmetric coupling to the electrodes ($\Gamma_1 = 0.02 = \Gamma_2$) and different values of charging: $U = 0$ (broken line), $U = 0.25$ (thin solid line) and $U = 0.5$ (thick solid line): (a) current-voltage $I(V) = -I(-V)$ and (b) conductance-voltage $G(V) = G(-V)$ characteristics. Temperature: $k_B T = 1/40$ eV.

### 3. Results and discussion

In order to fabricate molecular devices, different types of nanowires are possible to provide. Anyway, the most popular are organic molecules [1], where conduction is due to π-conjugated molecular orbitals. In particular, linear carbon-atom chains containing up to 20 atoms connected at the ends to metal atoms have been synthesized [19] and recognized as ideal one-dimensional wires [20]. As an example, we have studied linear chain containing 12 carbon atoms weakly connected at the ends to the metal electrodes. Since only π-electrons are involved into the conduction process, initial electronic structure of the molecule is obtained through the use of Hückel Hamiltonian (with orthogonal atomic basis set of states) in the procedure of diagonalization [12] ("bare" energy levels are given in eV: $\pm 0.603$, $\pm 1.773$, $\pm 2.840$, $\pm 3.743$, $\pm 4.427$, $\pm 4.855$). For further calculation we also assume realistically that Fermi level of unbiased electrodes is closer to the LUMO level ($\varepsilon_F = 0.3$) [21].

*3.1 Charging-induced effects*

Fig.1 presents transport characteristics of a 12-atom nanowire connected symmetrically to two metallic electrodes at room temperature. For the case of $U = 0$ our CB model exactly reproduces the results obtained with the help of generalized Breit-Wigner formula for $U = 0$ [12]. However, for $U > 0$ we observe the richer structure of such characteristics than in the case of SCF approach. In particular, some additional steps in the I-V dependences and some additional peaks in the conductance spectra are documented. This is a direct consequence of Coulomb interactions into the molecular system. The distance between these additional and main peaks in the G-V functions is $V = 2U/e$ and they appear in the direction outside the conductance gap (CG). Of course, for the case of symmetric coupling we obtain symmetric transport characteristics (there is no physical reason for different behavior).

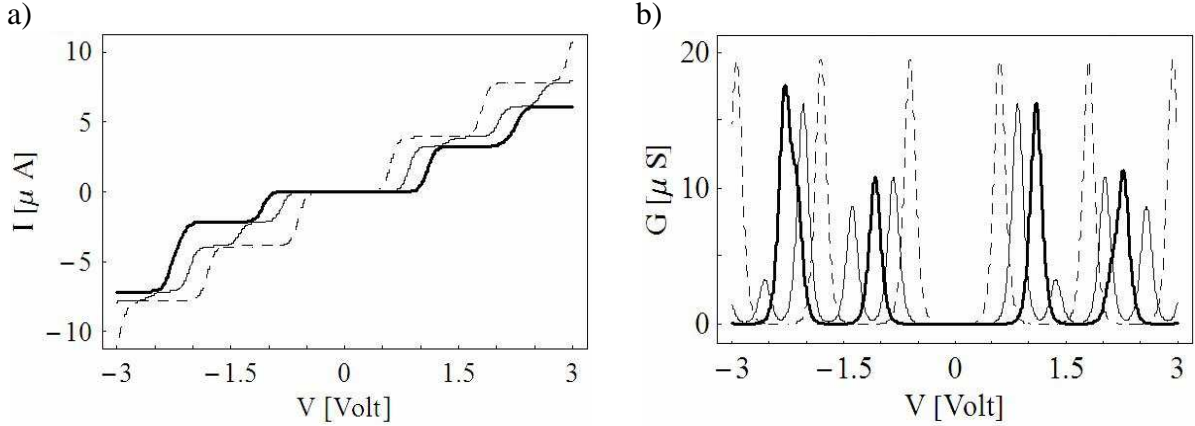

Figure 2: Results for linear chain of 12 carbon atoms for asymmetric coupling to the electrodes ($\Gamma_1 = 0.01 = 4\Gamma_2$) and different values of charging: $U = 0$ (broken line), $U = 0.25$ (thin solid line) and $U = 0.5$ (thick solid line): (a) current-voltage $I(V) \neq -I(-V)$ and (b) conductance-voltage $G(V) \neq G(-V)$ characteristics. Temperature: $k_B T = 1/40$ eV.

One of the most important results is associated with real suppression of the current at higher voltages (or equivalently partition of conductance peaks) due to Coulomb blockade, where jump-like character of I-V curve is still evident. In the SCF regime we observe smoothing of I-V curves due to Coulomb interactions, where suppression of the current is possible only for extremely large $U$-parameter (while current staircase is invisible in this case). This suppression is desirable effect from theoretical point of view as it can partly explain discrepancies between calculated and experimentally obtained values of the current (differences of two-three orders of magnitude) [21]. So far this divergence was closely connected to some coupling effects, such as: the atomic-scale contact geometry, the nature of molecule-to-electrodes coupling (chemisorption or physisorption) or even the changes of surface properties due to adsorption of molecular layer. However, here we show that also Coulomb blockade phenomenon can reduce the current at molecular scale. Among the other factors that can alter the value of the current flowing through the molecular junction one can also enumerate: some thermal effects (hot electrons and vibrational coupling) or local disorder in the electrodes near the contacts (electron localization).

On the other hand, charging enters the picture of conductance only at higher voltages, similarly as in the SCF regime. However, here we observe charging-induced increase of conductance gap for $U > 0$, where the mentioned gap strongly depends on the value of $U$-parameter. It should be also noted that it is also possible to reduce CG quantity, assuming that charging parameter is negative $U < 0$ (this choice can be justified by a strong electron-phonon interaction in the molecular quantum dot [22] or by certain pure chemical arguments as in the case of mixed valence complexes [23]). The total conductance gap in the CB regime is given through the relation: $4(|\varepsilon_F - \varepsilon_{LUMO}| - \Delta + U/2)$ (where: $\Delta \sim 4k_B T + \Gamma_1 + \Gamma_2$). Here we should remind the reader that in the SCF limit, CG quantity does not change upon charging [12,24], being expressed as follows: $4(|\varepsilon_F - \varepsilon_{LUMO}| - \Delta)$. Furthermore, it should be noted that some state-of-art first-principles calculations overestimate conductance gap in comparison with experimental data [21]. Here we indicate that Coulomb blockade phenomenon has a great influence on determination of CG quantity.

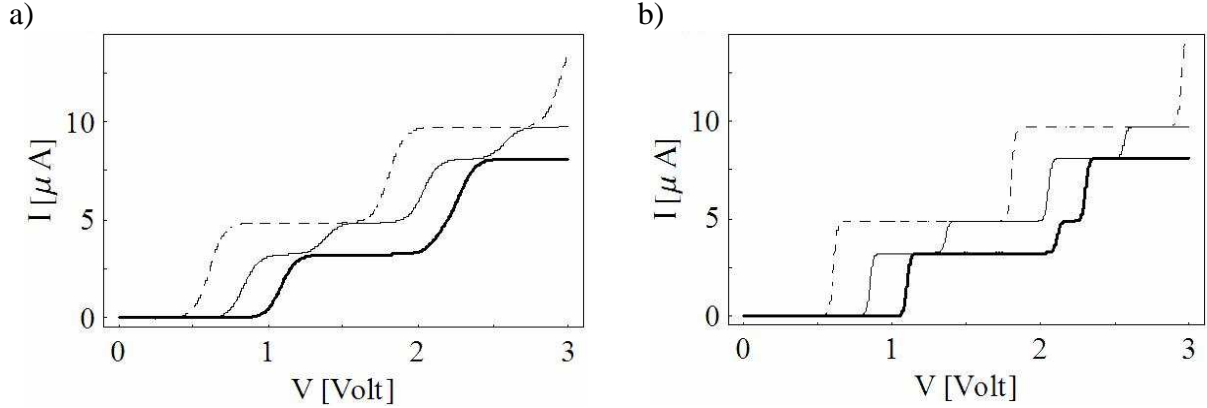

Figure 3: Current-voltage $I(V) = -I(-V)$ characteristics for linear chain of 12 carbon atoms for symmetric coupling to the electrodes ($\Gamma_1 = 0.02 = \Gamma_2$) and different values of charging: $U = 0$ (broken line), $U = 0.25$ (thin solid line) and $U = 0.5$ (thick solid line): (a) temperature $k_B T = 1/40$ eV, (b) temperature $k_B T = 1/170$ eV.

In Fig.2 we plot transport characteristics for a 12-atom nanowire connected asymmetrically to two metallic electrodes at room temperature. On the enclosed pictures we can see that there is no possibility to generate asymmetric transport dependence in the absence of Coulomb blockade (for $U = 0$), although the molecule-to-electrodes coupling is asymmetric. This conclusion is in good agreement with SCF results [12,24]. For the case of $U > 0$ we can observe charging-induced rectification effect, where the magnitude of the current depends on the polarity of applied bias (for higher voltages). Generally, the larger $U$ - parameter, the stronger asymmetry in I-V dependence, and the higher voltages should be applied in order to observe diode-like behaviour (since conductance gap is increased with increasing of charging energy). Such rectification effect has been demonstrated in a number of experiments involving monolayer and multilayer films as well as a single molecule type junctions and STM measurements [8,9]. Within our model, this behaviour is explained as a combined effect of asymmetry coupling to the reservoirs and Coulomb blockade phenomenon.

*3.2 Temperature-induced effects*

Fig.3 documents I-V dependences of a 12-atom nanowire connected symmetrically to two metallic electrodes at two different temperatures: room temperature ($k_B T = 1/40$ eV) and liquid nitrogen ($k_B T = 1/170$ eV), respectively. Significant temperature-induced smoothing of I-V curves (or equivalently broadening of conductance peaks) is observed, while this effect is not so essential in the case of SCF transport regime. Here closely-located current steps (or conductance peaks) can even merge due to the mentioned phenomenon (compare the curves for $U = 0.5$). Besides, it should be also remembered about temperature-generated shifting of resonances (current steps or conductance peaks) [17]. However, this effect seems to be negligibly small for MQD-based junctions within reasonable range of model parameters.

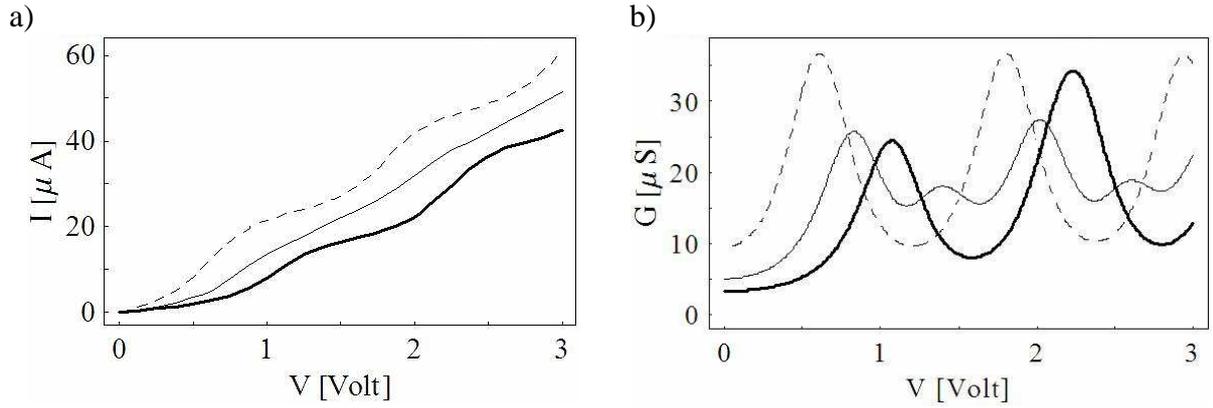

Figure 4: Results for linear chain of 12 carbon atoms for symmetric coupling to the electrodes ($\Gamma_1 = 0.1 = \Gamma_2$) and different values of charging: $U = 0$ (broken line), $U = 0.25$ (thin solid line) and $U = 0.5$ (thick solid line): (a) current-voltage $I(V) = -I(-V)$ and (b) conductance-voltage $G(V) = G(-V)$ characteristics. Temperature: $k_B T = 1/40$ eV.

*3.3 Broadening-induced effects*

In Fig.4 we show transport characteristics of a 12-atom nanowire connected symmetrically to two metallic electrodes at room temperature. But now let us analyze the case of stronger coupling, where broadening of energy levels can not be neglected (in our model this effect is taken into account by Lorentzian densities of states). As expected, with increasing the strength of the coupling to the electrodes we observe an increase of values of the current flowing through the device. Also significant broadening-induced smoothing of I-V curves (or equivalently extending of conductance peaks) is observed, similarly as in the case of SCF transport regime. However, we should keep in mind that this model is not well justified in the case of very strong coupling. Anyway, here we have shown how the broadening due to the coupling with reservoirs can be included into this formalism and the possible effects associated with this broadening.

**4. Conclusions**

In this work we have used simplified model based on rate-equation approach in order to study the qualitative changes of transport characteristics in molecular devices in the Coulomb blockade regime. We have found few interesting effects, such as: the suppression of the current at higher voltages, charging-induced rectification effect, charging-generated changes of conductance gaps and the overall structures of transport characteristics. In this work we have focused our attention especially on the differences and similarities between obtained results in the CB transport regime and those obtained in the SCF transport regime. Finally, temperature-induced as well as broadening-generated smoothing of current steps was also confirmed.

**Acknowledgements**

The author is very grateful to Gloria Platero for many interesting discussions.

## Appendix

Here we present standard rate-equation approach to transport via discrete energy states [7,14-17], which was used in section 2 in order to develop presented model. The time evolution of the occupation probability of arbitrary state $|\alpha\rangle$ is given through the so-called master equation:

$$\frac{dP_\alpha}{dt} = \sum_\beta \left( \gamma_{\beta \to \alpha} P_\beta - \gamma_{\alpha \to \beta} P_\alpha \right). \tag{A.1}$$

Here we are interested only in stationary state limit, where all the probabilities are time-independent $dP_\alpha / dt = 0$, hence:

$$\sum_\beta \gamma_{\beta \to \alpha} P_\beta = \sum_\beta \gamma_{\alpha \to \beta} P_\alpha. \tag{A.2}$$

Moreover, the summation over all the probabilities must be equal to one (normalization):

$$\sum_\alpha P_\alpha = 1. \tag{A.3}$$

Once one has solved the master equation for the individual probabilities, the current flowing through the device can be calculated from the following formula:

$$I_q = e \sum_{\alpha,\beta} (\pm) \gamma^q_{\alpha \to \beta} P_\alpha, \tag{A.4}$$

where $q = L, R$ denotes the contact (left and right, respectively) and the sign of the $q$-term of transition rate $\gamma^q_{\alpha \to \beta}$ depends on whether the transition $\alpha \to \beta$ gives a positive (plus) or negative (minus) contribution to the current.